\documentstyle[aas2pp4]{article}

\def\msun{{\,M_\odot}}

\def\refindent{\par\noindent\hangindent=3pc\hangafter=1 }

\def\apj#1#2#3{\refindent#1, {\it ApJ}, {\bf#2}, #3.}
\def\apjlett#1#2#3{\refindent#1, {\it ApJ (Letters)}, {\bf #2}, #3.}

\def\mnras#1#2#3{\refindent#1, {\it MNRAS}, {\bf#2}, #3.}

\def\cm{{\rm\,cm}}

\def\sec{{\rm\,s}}

\def\gm{{\rm\,g}}

\def\kelvin{{\rm\,K}}
\def\erg{{\rm\,erg}}

\def\>{$>$}
\def\<{$<$}

\def\simlt{\lower.5ex\hbox{$\; \buildrel < \over \sim \;$}}
\def\simgt{\lower.5ex\hbox{$\; \buildrel > \over \sim \;$}}
\def\sqr#1#2{{\vcenter{\hrule height.#2pt
      \hbox{\vrule width.#2pt height#1pt \kern#1pt
         \vrule width.#2pt}
      \hrule height.#2pt}}}

\begin{document}

\centerline{Accepted for publication in the Astrophysical Journal (Letters)}
\bigskip
\title{Big Blue Bump and Transient Active\\
    Regions in Seyfert Galaxies}

\author{Sergei Nayakshin and Fulvio Melia\altaffilmark{1}}
\affil{Physics Department and Steward Observatory, University of Arizona,
    Tucson, AZ 85721}

\centerline{Revised April 24, 1997}




\altaffiltext{1}{Presidential Young Investigator.}


\begin{abstract}
An important feature of the EUV spectrum (known as the Big Blue
Bump, hereafter BBB) in Seyfert Galaxies is the narrow
range in its cutoff energy $E_c$ from source to source, even though
the luminosity changes by 4 orders of magnitude.  Here we show
that if the BBB is due to accretion disk emission, then in order 
to account for this ``universality'' in the value of $E_c$, the 
emission mechanism is probably optically thin bremsstrahlung. In addition,
we demonstrate that the two-phase model with active regions localized
on the surface of the cold disk is consistent with this constraint
if the active regions are very compact and are highly {\it transient},
i.e., they evolve faster than one dynamical time scale.
\end{abstract}




%

\section{Introduction}

The UV to soft X-ray spectrum of many Active Galactic Nuclei (AGNs) may
be decomposed into a non-thermal power-law component and the so-called 
Big Blue Bump (BBB), which cuts off below about 0.6 keV (e.g., 
Sanders et al. 1989).  A major obstacle in constraining the characteristics
of the BBB has been that it lies in the difficult to observe EUV and very 
soft X-ray region.  In recent years, however, there has been considerable 
progress in this direction (e.g., Walter \& Fink 1993; Walter et al. 1994; 
Zhou et al. 1997). In particular, Walter \& Fink (1993) observed that 
the spectral shape of the bump component in Seyfert 1's hardly varies, 
even though the luminosity $L$ ranges over 6 orders of magnitude from source
to source.  Soon after, Walter et al. (1994) concluded that the
cutoff energy $E_c$ of the BBB (when fitted as a power-law with
exponential rollover) was very similar in different sources
whose luminosities varied by a factor of 10$^4$. Although the data
were not precise enough to distinguish between different emission
mechanisms, Walter et al. (1994) pointed out that 
if the variations in the ratio of the soft X-ray excess to UV flux
from one object to another are interpreted as a change in the temperature 
of the BBB, then this change is smaller than a factor of 2.
Confirming conclusions follow from the work of Zhou et al. (1997). 

Early theoretical work on the BBB spectrum focused on the role of optically 
thick emission from the hypothesized accretion disk surrounding the central 
engine (e.g., Shields 1978; Malkan \& Sargent 1982;  Czerny \& Elvis 1987;
Laor \& Netzer 1989).  However, this mechanism is now facing several obstacles
(e.g., Barvainis 1993; Mushotzky et al. 1993).  An alternative model, 
in which the BBB is interpreted
as thermal, optically thin free-free radiation, has been proposed by Antonucci 
\& Barvainis (1988), Barvainis \& Antonnuci (1990), Ferland et al. (1990), and
Barvainis (1993).

In parallel to this, several authors have also been concerned with the
spectrum produced by X-ray irradiated disks (e.g., Ross \& Fabian 1993; 
Zycki et al. 1994; Collin-Souffrin et al. 1996; Sincell \& Krolik 1997).  
In these studies, the incident X-ray intensity is always assumed 
to be stationary in time. However, the most recent work on the 
physics of the high-energy sources
suggests that a likely origin for the illuminating X-rays are magnetic flares 
above the surface of the cold accretion disk (e.g., Haardt et al. 1994;
Nayakshin \& Melia 1997).  The lifetime of these flares is probably much
smaller than the disk's hydrostatic time scale, for which the assumption of
time-independent reflection is in that case not warranted. 

One would hope that in a self-consistent picture, this locally time-dependent
X-ray heating of the cold disk is linked to the BBB emission.
In this {\it Letter}, we attempt to identify which of the various models for
the BBB emission can account for the observed near-independence of $E_c$ on
the AGN luminosity.  Our main goal is to determine if the viable mechanism
can arise as a result of X-ray illumination of the disk by a {\it transient 
magnetic flare}.  In so doing, we show that the structure of the time-dependent 
reflecting layer is very different from that assumed in the time-independent 
case.  We will show that short-lived flares on the surface of the accretion
disk may allow us to resolve the apparent inconsistency between the constancy
of $E_c$ and the wide range of luminosities, and we will demonstrate that
the expected BBB temperature compares favorably with the value observed
in these sources.

\section{The Radiation Flux and Emission Mechanisms}

AGNs are thought to accrete both from their nearby environments via
the Bondi-Hoyle process and from the tidal disruption of stars, though
over time, the former is dominant (e.g., Melia 1994).  At least initially,
the accretion rate is therefore $\dot M\sim M^2$, where $M$ is the
black hole mass (e.g., Shapiro \& Teukolsky 1983), but this constitutes 
a runaway process in the sense that $L/L_{Edd}\propto t$, where $t$ is the 
time, and $L_{Edd}$ is the Eddington luminosity.  When $L\rightarrow L_{Edd}$, 
the outward radiation pressure presumably suppresses the inflow, with the 
effect that $L$ saturates at the value $\sim L_{Edd}\propto M$.  A second
argument in favor of the supposition that the ratio $L/L_{Edd}$ is probably 
independent of $M$ is the fact that we observe very similar X-ray spectra
for objects of very different luminosities (e.g., Zdziarski et al.
1997), for otherwise the disk structure would differ from source to source,
giving rise to different spectra.
As a statistical average, we thus expect that $L\propto M$.

In view of this, let us next examine how the various different emission
mechanisms fare in their prediction of the BBB cutoff energy $E_c(L)$.
For any radiation process, the flux $F$ scales as $L$ over the
emitting area, which itself scales as $M^2$.  Thus, in general we expect that
\begin{equation}
F\sim L^{-1}. 
\end{equation}
The blackbody flux is $F_{\rm bb}=\sigma T^4$, where $T$ is the effective 
temperature,
and so $T\sim L^{-1/4}$.  Thus, when the luminosity varies by 4 orders of
magnitude, it is expected that the blackbody temperature ought to itself
vary by a factor of 10.  This is not consistent with the observations
discussed above.

A more sophisticated treatment of the disk structure in its inner
region shows that the scattering opacity can dominate over the absorptive one,
and thus the emission spectrum will differ from the blackbody spectrum.
Even for a radiation-dominated configuration,
the disk is likely to be effectively optically thick (see Eq. 14 of 
Svensson \& Zdziarski 1994, for $r=7$, $f\leq 1/2$ and $\alpha\sim 0.1$). 
The disk may therefore radiate as a `modified blackbody' (Rybicki \&
Lightman 1979), for which the flux is then given by 
\begin{equation}
F_{\rm mb}\sim 2.3 \times 10^7 T^{9/4}\rho_d^{1/2}
\erg\;\cm^{-2}\;\sec^{-1}{\rm ,}
\end{equation}
where $\rho_d$ (in $\gm\;\cm^{-3}$) is the disk mass density and $T$ is 
in Kelvins.  For the likely situation of a radiation-dominated disk, 
$\rho_d\sim L^{-1}$, and so $T\sim L^{-2/9}$, which again is not
consistent with the data.

Optically thin bremsstrahlung, on the other hand, produces a flux
\begin{equation}
F_{\rm ff}= \varepsilon_{\rm ff}\, d =
6.1\times 10^{20} T^{1/2} \rho^2 d\,
\erg\;\cm^{-2}\;\sec^{-1} {\rm ,}
\end{equation}
where $\varepsilon_{\rm ff}$ (erg$\cm^{-3}\sec^{-1}$) is the free-free 
emissivity (Rybicki \& Lightman 1979), and $d$ is the geometrical 
thickness of the emitting region.  This expression assumes that the
ions are protons and that their density is equal to that of the
electrons.  Thus, since $d$ presumably scales as $R_g\equiv
2GM/c^2\propto L$, $T$ is independent of $L$ because of Equation (1).
We note, however, that because of the very weak dependence of 
$F_{\rm ff}$ on $T$, the actual temperature is likely to be constrained
by the local physics of the emitting region rather than by the integrated
flux (as would be the case for the other two emission mechanisms).
It remains to be seen if a detailed study confirms this heuristic argument. 

\section{Time-Independent X-ray Illumination of the Disk}

It is well known that an active region (AR) radiating X-rays above
the cold disk will produce a reflected component and a reprocessed
UV spectrum due to the absorbed X-ray flux (Guilbert \& Rees 1988; 
Lightman \& White 1988; White et al. 1988).  The characteristic Thomson 
optical depth $\tau_T$ at which the incident X-rays are absorbed or
scattered to lower energies is of order a few.  It is this layer,
sometimes referred to as the X-ray skin, that re-emits the deposited
energy.

How this energy is re-radiated depends critically on the 
photoionization of the reflecting layer by the incident X-ray flux.  
This formidable task has been addressed by, e.g., Ross \& Fabian 
(1993), Zycki et 
al. (1994), Czerny \& Zycki (1994), and Sincell \& Krolik (1997).  
Czerny \& Zycki (1994) suggested that the ``universal'' shape of the 
BBB is explained by the atomic physics of the reflection process. 
They fitted the spectra of several Seyfert 1 Galaxies and concluded 
that in almost all of them, the soft X-ray excess (which we consider 
to be a part of the BBB) is well represented by reflection/reprocessing 
of the incident X-ray flux.  However, their required normalization 
of the reflected component relative to the direct one was as large 
as 2-3, which is not consistent with the geometry of a flat 
reflecting disk.  Instead, the normalization in this case is expected 
to be unity, as confirmed from fits of the reflected X-ray spectrum
in an almost neutral absorber (e.g., Zdziarski et al. 1997). In addition, 
their assumed UV flux was larger than the X-ray flux by a factor of 
30-100, whereas the observed ratio is closer to a few.  
They also assumed a fixed UV-temperature,
which is rather ad hoc.  As far as the UV portion of the spectrum 
is concerned, the photoionized reflection models produce a temperature 
that is either well below (e.g., Sincell \& Krolik 1997) the observed 
value $\sim 60$ eV, or one that is strongly dependent on $M$
(for example, $T$ changes by about a factor of 2 for a change in
$M$ by a factor of 10 in Figure 2 of Ross \& Fabian 1993). 
Physically, this is explained by the fact that in the stationary case 
the thermal equilibrium in the whole disk below the X-ray emitting region is
established, and thus the characteristic temperature of the UV emission
is representative of the disk itself rather than the reflection process.
The arguments given in \S 2 then show that the undesirable correlation between
$T$ and $L$ ensues.

We also note that calculations of the static X-ray reflection/reprocessing
cannot be simply extended to the reflection of X-rays
from such dynamic processes as magnetic flares, whose presence seem to be
necessary in order to explain the hard X-ray spectrum of Seyferts
(e.g., Nayakshin \& Melia 1997).
For the static description to be applicable, thermal (in addition
to hydrostatic) equilibrium must be established. This means that the 
X-ray flux would need to be static on a disk thermal time scale, which 
is $\sim 1/\alpha$ times longer than the hydrostatic time scale 
$t_h$. However, the shearing time in the disk is of order $t_h$, 
so that any magnetic structure (responsible for the X-ray flux) 
should decay away during that time. In addition, as we shall see
below, stationary X-ray reflection predicts an ionization state
that is too high for the reflector, while observations of Seyfert galaxies
require a nearly neutral absorber (Zdziarski et al. 1997).
On the basis of these deficiencies, we suggest that {\it stationary} 
X-ray illumination is probably not a viable mechanism for producing
X-ray reflection/reprocessing indicated by observations of Seyfert Galaxies.

\section{Time-Dependent X-ray Illumination of the Disk}

Fortunately, many of the desirable features of the X-ray reprocessing
inferred from observations will apply also for time-dependent
X-ray illumination of the disk, which we now consider. In the following, the principal
distinction between this and the time-independent case is the type of equilibrium
established during a flare. The X-ray skin, being a very small fraction of the
total thickness of the disk will adjust very quickly to a quasi-equilibrium
with the incident X-radiation. The rest of the material below the skin 
will be out of equilibrium due to the fact that the time scale required to 
establish such a state is far longer than the flare lifetime.

The compactness $l$ of the AR (here
assumed to be where the magnetic flare occurs) is defined according to
$l\equiv F_{\rm x}\sigma_T \Delta R_a/m_e c^3$, where $\Delta R_a$ is the
typical AR size.  The value of $l$ is expected to be rather high
($\sim 100$, Zdziarski et al. 1997).  The incident X-ray flux $F_{\rm x}$ 
can be deduced from $l$ and $\Delta R_a$: 
\begin{equation}
F_{\rm x} \equiv {l m_e c^3\over \Delta R_a \sigma_T} =
3.6\times 10^{17} {\rm erg}\, {\rm cm}^{-2} {\rm sec}^{-1}
{l_2\over \Delta R_{13}}\;,
\end{equation}
where $\sigma_T$ is the Thomson cross section, $l_2\equiv l/100$
and $\Delta R_{13}\equiv \Delta R_a/10^{13}{\rm cm}$.  The scaling
of $\Delta R_a$ is based on the expectation that it should be of 
order the accretion disk scale height ($H_d$). 

It is not difficult to see that when this flux turns on, the radiation 
ram pressure on the surface of the disk greatly exceeds the equilibrium 
thermal pressure from within.  The X-ray skin therefore gains momentum 
and an inward plow phase is initiated that very quickly slows down
the density wave.  Using conservation of momentum, we estimate
the velocity of the inwardly driven gas to be $v\sim \sqrt{F_{\rm x}/
c\rho_{\rm d}}$, where $\rho_d$ ($\approx 10^{-10}$ g cm$^{-3}$;
Svensson \& Zdziarski 1994) is the density in the disk.   For these
scaled parameter values, $v\lesssim 10^9$ cm s$^{-1}\,\ll c$.  The density 
itself is expected to increase in the X-ray skin until the reprocessed
UV free-free emissivity balances the incident X-ray flux, at which point,
\begin{equation}
\rho = 2.8\times 10^{-7}\gm\cm^{-3}
{l_2\over \Delta R_{13}} T_5^{-1/2} \left (\tau_{\rm x}/3\right)^{-1}
{\rm ,}
\end{equation}
where $T_5\equiv T/10^5$ K. (We here used Equation (3) for the free-free 
emissivity.) If the X-ray skin were to contract further,
the UV emissivity would exceed the incident radiation flux which clearly
violates energy conservation.  Comparing this with the gas density $\rho_{\rm d}$
in a cold accretion disk, we see that the latter is smaller than 
$\rho$ by 2-3 orders of magnitude. Although the scattering 
optical depth of the X-ray skin 
is $\sim$ few, the Compton $y$-parameter describing the importance of the
Comptonization (Rybicki \& Lightman 1979) 
is $\lesssim 10^{-2}$, and thus Comptonization 
of UV photons in the X-ray skin is unimportant. 
Moreover, the absorption optical depth of the X-ray skin is negligible 
for the densities
considered here, and the resulting spectrum is that of optically 
thin free-free emission (see below for a discussion of the free-bound emission).

This argument does not yet fix $T$, which we simply
scaled to the value $10^5$ K in the above equation.  With two unknowns,
$\rho$ and $T$, we need a second physical constraint, which we take to
be the equilibrium between internal pressure $2 (\rho/m_p) kT$ of the 
compressed gas and the incident X-ray flux. Note that we can, indeed, 
assume the X-ray skin to be in quasi-equilibrium: the time scale
for establishing this equilibrium is much smaller than the thermal
disk time scale, and thus it can be shown that 
even though the underlying disk does not
reach equilibrium during the short lived flare, the much thinner 
X-ray skin does. The exact pressure equilibrium during a flare
could only be found in a full-scale, self consistent calculation,
where the dynamics of the gas immediately below the X-ray skin is
taken into account.  We shall not attempt to carry this out here, but 
rather rely on some simple estimates. For simplicity, we
calculate the radiation force in the Eddington approximation. 
The radiation energy flux is $F_e = - c/3\; (du_{\rm rad}/d\tau)$, 
where $u_{\rm rad}$ is the radiation energy density and $\tau$ is the 
total optical depth (Rybicki \& Lightman 1979).  To find the total 
compressional force $F_{\rm rad}$ acting on the absorbing/reflecting 
layer, we integrate over $\tau$:
\begin{equation}
F_{\rm rad} = 1/c\,\int_{0}^{\tau_{\rm x}}d\tau\; F_e = \onethird\,
\left [u_{\rm rad}(0) - u_{\rm rad}(\tau_{\rm x})\right ]\;.
\end{equation}
We emphasize that there is no net flux in the time-independent case
if all the energy is dissipated in the corona (e.g., Sincell \& 
Krolik 1997). If the energy dissipated within the cold disk is not zero,
then there is a net outward flux balanced by gravity.
In the case of a magnetic flare (i.e., a time-dependent situation), 
however, the net radiative flux points {\it toward} the disk, so the 
radiation pressure squeezes the gas rather than expands it. 

In a steady state situation, we expect $F_{\rm rad}=0$. At the other
extreme, when all the incident X-ray flux is re-radiated back to
the corona, $u_{\rm rad}(\tau_{\rm x}) = 0$ and $u_{\rm rad}(0) =
2\sqrt{3} F_{\rm x}/c$ (the skin is then just a mirror reflecting
the incident momentum flux).
The pressure equilibrium condition is therefore
\begin{equation}
2 (\rho/m_p) k_b T = 2/3^{1/2}\, A \left[ F_{\rm x}/c\right ]\;,
\end{equation}
where $k_b$ is Boltzmann's constant, and the unknown parameter
$0<A<1$ reflects our ignorance of the specific details in this
layer (note that radiation pressure from the intrinsic disk emission
is negligible compared with the value found from Equation [7]
in the context of the two-phase models, see below).
The X-ray flux drops out of the equation when we include
the energy balance condition $F_{\rm x} = \varepsilon_{\rm ff} d = 
F_{\rm UV}$, in which $F_{\rm UV}$ is given by Equation (3). 
It is this cancellation of the exact value of $F_{\rm x}$ 
that leads to the mass-invariance of the temperature.
Assuming we know the exact value of $A$ (our guess is that it
is probably between $1/2$ and 1), we therefore infer
a unique value for $T$: 
\begin{equation}
T = 3.8 \times 10^5 \kelvin\left (A\tau_{\rm x}/\sqrt{3}\right )^2\;.
\end{equation}

The validity of this treatment rests on the assumption that the intrinsic
disk flux is negligible compared to the local X-ray flux from the active
region, for otherwise the compressional effects will not work to produce
the required UV flux and BBB temperature.  It is straightforward to see
(using the results of Svensson \& Zdziarski 1994, for example)
that this condition is certainly met when $l\gg 1$, unless 
$\Delta R_a \gg H_d$, which is highly unlikely for a magnetic flare.
It is well known that the condition $l\gg 1$ is a key ingredient in 
two-phase corona-accretion disk models (Svensson 1996; Zdziarski et 
al. 1997; Nayakshin \& Melia 1997), so it is automatically satisfied in
situations where the two-phase model is valid.

We have not included the free-bound emissivity in our simple calculations.
We intend to carry out a more careful simulation in the future,
but we expect that this will only strengthen our case due to
the following reasons. The free-bound emissivity is a very strongly
decreasing function of temperature in the domain $T\sim 10^5$ - 
few $\times 10^6$ K. Let us approximate this situation by defining a 
critical temperature $T_c$, such that below $T_c$ the emission is dominated 
by the free-bound process, and that it is dominated by the free-free 
emission above it. Some preliminary calculations show that $T_c\sim 4\times 10^5$ K
(Kallman 1997). In the context of our simple calculations here,
to take into account the free-bound emissivity in addition to the 
free-free emissivity, the temperature derived in Equation (8)
should be multiplied by the factor $[1+\varepsilon_{fb}/\varepsilon_{ff}]^2$,
where $\varepsilon_{fb}$ and $\varepsilon_{ff}$ are the free-bound and the 
free-free emissivities, respectively. Above
$T_c$ the free-bound correction is negligible and equation (8) applies.
Below $T_c$ Equation (8) is corrected by the large factor, and leads to
temperatures that are very likely to be larger than $T_c$. Physically this simply
means that below $T_c$ a more careful calculation will lead to higher temperature
than predicted by Equation (8), thus making it closer to $T_c$.
In other words, temperatures much different than $T_c$ are unlikely to
be achieved in the X-ray skin during the flare. Since $T_c$ is of the same
order as the temperature given in Equation (8), we expect that 
a careful addition of the free-bound emission to our calculations
will preserve our conclusions with a possible 
correction to the value of the temperature derived in this paper.
In particular, independence of the temperature of the reprocessed
emission on the luminosity will remain a distinguishing feature of the
model.

Let us now estimate the ionization parameter $\xi$ (e.g., Ross \& Fabian 1993)
in the X-ray skin. Zycki et al. (1994) have shown that the quality of the
X-ray data are high enough to distinguish between reflection from 
weakly ionized/neutral gas and ionized gas if the latter has 
ionization parameter $\xi$ larger than about 200. Since the data are
adequately fitted with a neutral reflector, we then require that the ionization
parameter of the gas should not exceed 200. Since the gas density in 
the X-ray skin is higher than in the static models, our estimated value for $\xi$
in the X-ray skin is always $\lesssim $ few tens, and thus is
consistent with observations. On the other hand, static reflection 
is characterized by a gas density smaller by typically 2-3 orders of magnitude,
and therefore the ionization parameter is too high for this reflection. 
The situation can be improved if the X-ray emitting region covers
the whole disk (static corona)
and has a lower X-ray flux on average. However, such models
are inadequate for Seyfert galaxies on the basis of the observed
UV/X-ray energy partitioning (e.g., Svensson 1996) and should be rejected.

\section{Discussion and Conclusions}

We have discussed some of the issues pertaining to the physics of 
magnetic flares above the surface of an accretion disk in Seyfert
galaxies, and we have produced a simple, yet physically consistent, 
explanation for the relatively universal value of the BBB temperature
in the face of large variations in luminosity.  The UV emitting layer 
settles down to a unique temperature given by Equation (8) (with possible
corrections due to the free-bound emission to be computed in future work),
independently of the incident X-ray flux $F_{\rm x}$.  In effect, 
one could define a so-called ``bremsstrahlung temperature'' $T_{\rm brems}$
as the temperature at which the radiation pressure of the bremsstrahlung 
photons in an optically thin plasma is equal to the gas pressure. 
Equation (8) gives a value of $T$ that is close to $T_{\rm brems}$.
If this interpretation of the BBB is correct, we then see that the physical state
of the re-emitting layer arises from the adjustment of the gas density 
to provide a sufficiently large emissivity to balance the energy
deposition of the incident X-rays, whereas the temperature adjusts in
order to re-establish pressure balance. 

The key feature that distinguishes the time-dependent X-ray 
absorption/reflection model from the static one is the transient, inverse 
density gradient produced in the disk---the incident X-ray flux
compresses the upper layers, whereas the particle density profile
in a static configuration is either constant with height when radiation
pressure dominates (Shakura \& Sunyaev 1973), or is Gaussian 
if no internal dissipation in the disk is assumed (e.g., Sincell 
\& Krolik 1997).   In the latter two cases,
the emission mechanism is either a modified blackbody,
or a true blackbody, depending on how much accretion power is directly
released within the cold disk (Sincell \& Krolik 1997), but in both
situations the temperature of the reprocessed emission is representative 
of the disk properties rather than the X-ray skin properties, and this 
leads to a strong dependence of the observed 
BBB temperature on the source luminosity, which appears to be in
conflict with the data. The static X-ray reflection is also
physically ill-suited for describing the highly dynamic 
reflection of X-ray emission from magnetic flares.

Our derived temperature compares favorably with
the cutoff energy range ($\sim$ 36-80 eV) observed in the sample
of Zhou et al.(1997), with the exception of NGC 4051. This exception
is probably explained by assuming that it is intrinsically much less massive
than a typical AGN ($\sim 10^8 \msun$). Equation (3)
requires a density that scales as $M^{-1}$, and so for this
source, one obtains a free-free optical depth for the X-ray skin 
comparable to or larger than unity. The emission
mechanism is then blackbody rather than bremsstrahlung, 
and thus the BBB temperature scales with $M$ for masses lower 
than some critical value ($\sim$ few $\times 10^5 \msun$). 
We are pursuing this aspect of the problem in more detail and we shall 
report the results in future publications.

A magnetic flare irradiates not only the region directly below it,
but also regions further away where the X-ray flux is lower, and
which therefore probably results in a more complicated spectrum than
we have considered here.  It is also possible that the heated regions
continue to cool after the flare has subsided.  We anticipate that
the UV continuum may be a combination of all of these effects, on
top of the intrinsic disk emission.  Future observations of the
UV to the soft X-ray portion of Seyfert spectra may thus provide a
means of disentangling the various emission processes and enable us
to make more reliable comparisons between different models.

\section{Acknowledgments}

This work was partially supported by NASA grant NAG 5-3075.
We thank T. Kallman for carrying out an ionization
balance calculation that allowed them to estimate the importance of the 
free-bound emission relative to the free-free emission for the conditions
similar to the ones assumed in this paper.

%
%
%

{}

\end{document}